\documentclass[12pt]{iopart}
\usepackage{iopams}
\usepackage{graphicx}

\newcommand{\eqref}[1]{(\ref{#1})}
\begin{document}

\title[Wakefield acceleration in atmospheric plasmas]{Wakefield acceleration in atmospheric plasmas: a possible source of MeV electrons}

\author{M. Array\'as$^1$, D. Cubero$^2$, R. Seviour$^3$, J. L. Trueba$^1$} 

\address{$^1$ \'Area de Electromagnetismo, Universidad Rey Juan
Carlos, \\Camino del Molino s/n, 28943 Fuenlabrada, Madrid, Spain}
\address{$^2$ Dept. de F\'{\i}sica Aplicada I, Universidad de Sevilla, \\ Virgen de \'Africa 7, 41011 Seville, Spain}
\address{$^3$ University of Huddersfield, Huddersfield, UK}
\ead{manuel.arrayas@urjc.es}

\begin{abstract}
Intense electromagnetic pulses interacting with a plasma can create a wake of plasma oscillations. Electrons trapped in such oscillations can be accelerated under certain conditions to very high energies. We study the conditions for the wakefield acceleration to produce MeV electrons in atmospheric plasmas. This mechanism may explain the origin of MeV or runaway electrons needed in the current theories for the production of Terrestrial Gamma ray Flashes.  
\end{abstract}

\submitto{\JPD}
\maketitle

\section{Introduction}
Electromagnetic pulses interacting with a plasma can create a wake of plasma oscillations through the action of the nonlinear ponderomotive force \cite{dawson}. Electrons trapped in the wake can be accelerated to high energies. In the laboratory very high-power electromagnetic radiation from lasers is used to accelerate electrons to high energies in a short distance. However, in atmospheric plasmas the distance range goes from metres to kilometres, and the conditions for the wakefield acceleration takes place are different as we will make clear in this work. 

The acceleration of the electrons in atmospheres has been the subject of attention since the discovered of Terrestrial Gamma Flashes (TGF's). TGF's were first discovered by accident in 1994 \cite{ref1} and only limited experimental data is available. One of the goals of the European Space Agency (ESA) mission ASIM \cite{web} will be to provide experimental data on TGF's. Satellite measurements suggest that TGF's typically last from a few tens of microseconds to a few milliseconds ($50 \, \mu \mbox{s}$ -- $1 \, \mbox{ms}$) \cite{ref1,ref9,ref10,ref11}, exhibit an energy spectra extending from 1 MeV up to 100 MeV \cite{ref1,ref2, ref2p,ref3,ref12}, over a half conic-angle of $30^{\circ} - 40^{\circ}$ \cite{ref17,ref18} and have a fluence of 0.1-0.7 photons/cm$^2$. They seems to appear around {$15 \, \mbox{km}$} near the upper regions of thunderclouds \cite{ref5,ref6,armenia}. Any theory about the TGF generation should be able to successfully explain the observed duration, energy spectra, and photon fluence. There is some agreement that the emission is produced via Bremsstrahlung \cite{ref7,ref8} when high energy electrons collide with nuclei in the air releasing  energy. The satisfactory theory for TGF's needs to explain the origin of the high-energy electrons \cite{ref13} which can trigger the ignition of the gamma flashes. One  explanation is given by the feed-back theory \cite{ref2} in which seed electrons are provided by cosmic rays. Other possibilities are leader/streamer theory \cite{ref15}, in which seed electrons are produced in the streamer head and accelerated in the stepped leader electric field, and the current pulse version \cite{ref16}. Here we present an alternative scenario where those energetic electrons come from the interaction of a electromagnetic pulse with an existing plasma previously created. Although the idea of TGF's production by an electromagnetic pulse due to a lightning return stroke was explored previously \cite{lehtinen}, the novelty of our proposal is the interaction of the pulse with a plasma already present in the atmosphere and created for example by another discharge or any electromagnetic activity. Since the interaccion of the electromagnetic pulse is with a plasma, under certains conditions the braking in the motion of the electrons by friction does not play a role, as we will show in this work.   

 The outline of the paper is the following. First we revise how an electromagnetic wave can create a wakefield in a plasma. We study the conditions under which the acceleration of electrons to MeV energies is possible in the presence of atmospheric plasmas. The acceleration mechanism is demonstrated through computer simulations. We present the scenario for the production of the plasma and the electromagnetic pulse in  planetary atmospheres, and in particular for the Earth atmosphere we give some predictions. Finally we  summarise the main results and end with some conclusions.  



\section{Conditions for electromagnetic propagation of waves into a plasma}
Here we study the interaction of an electromagnetic wave with a plasma. Let us consider how a plasma modifies the propagation of electromagnetic waves \cite{ref19}. The starting point to describe the evolution of a plasma is the Vlasov equation, which up to first order in the expansion parameter $1/N_D$, reads
\begin{equation}
\label{fj}
\frac{\partial {f}_{j}}{\partial t}+{\bf v}\cdot \frac{\partial {f}_{j}}{\partial {\bf r}}+\frac{q_{j}}{m_{j}} \left( {\bf E} + {\bf v}_{j} \times {\bf B} \right)\cdot \frac{\partial {f}_{j}}{\partial {\bf r}} = \sum_k \left(\frac{\partial {f}_{jk}}{\partial t}\right)_C,
\end{equation}
being $f_j({\bf r},{\bf v},t)$ the distribution function of the $j$ species. The right hand side of \eqref{fj} represents the collisional terms. The parameter $N_D$ is the number of particles in the Debye sphere. For electrons, $N_D=4\pi n_e\lambda_{De}^3/3$, where $n_e$ is the electron density and $\lambda_{De}$ the electron Debye length. An useful expression of the Debye length under equilibrium conditions is given by $\lambda_{De}=69\sqrt{T/n_e}\, (\mbox{m})$, where $T$ is the equilibrium temperature of the electrons in Kelvin and $n_e$ is expressed in m$^{-3}$ \cite{spitzer}, so we have the condition
\begin{equation}
\label{Nd}
  N_D \approx 1.38 \times 10^6\, T^{3/2}/n_e^{1/2} \gg 1,
\end{equation}
for \eqref{fj} to be valid.
We recall that the condition to have a plasma is $\lambda_{De}\ll L$, being  $L$ the characteristic dimension of the system.  

From \eqref{fj} it is a standard procedure to calculate the first moment equations. We will assume that collisions do not change the number of species, so in the averaging we will take 
\[\int d{\bf v} \sum_k \left(\frac{\partial {f}_{jk}}{\partial t}\right)_C =0.\]
Hence, the change of the momentum of the $j$ species becomes
\begin{equation}
n_{j} \frac{\partial {\bf u}_{j}}{\partial t} + n_{j} {\bf u}_{j} \cdot \frac{\partial {\bf u}_{j}}{\partial {\bf r}} = \frac{q_{j}}{m_{j}} n_{j} \left( {\bf E} + {\bf u}_{j} \times {\bf B} \right) - \frac{1}{m_{j}} \frac{\partial p_{j}}{\partial {\bf r}} - \sum_{k \neq j} \left( \frac{\partial}{\partial t} n_{j} {\bf u}_{j} \right)_{k},
\label{collisioneq}
\end{equation}
where $n_j$ is the density and ${\bf u}_{j}$ the velocity of the $j$ species and $p_j$ the pressure. 

Let us consider a plasma composed of ions with positive charged $+eZ$ and electrons. We will assumed that ions form a fixed background with density $n_i=n_e/Z$, where $n_e$ is the corresponding electron density. Then we only need to treat the dynamics of the electron fluid. We can investigate the damping of an electromagnetic wave of the form ${\bf E}({\bf r})\exp(-i\omega t)$ considering the linearised plasma response.   Writing 
\[ \left( \frac{\partial}{\partial t} n_{e} {\bf u}_{e} \right)_{i}= \nu_{ei}n_e {\bf u}_{e}, \]
where $\nu_{ei}$ is the collisional frequency of the scattering of electrons by ions, from \eqref{collisioneq} to first order we get 
\begin{equation}
\frac{\partial {\bf u}_{e}}{\partial t} = -\frac{e}{m_{e}} \, {\bf E} ({\bf r}) \, e^{-i \omega t} - \nu_{ei} {\bf u}_{e}.
\label{collisionlin}
\end{equation}
The solution of \eqref{collisionlin} results
\begin{equation}
{\bf u}_{e} ({\bf r}, t) = \frac{-i e}{m_{e} (\omega + i \nu_{ei} )} \, {\bf E} ({\bf r})e^{-i \omega t} .
\label{colvel}
\end{equation}
The plasma conductivity $\sigma$ can be calculated from the current density of the plasma ${\bf j} = -e n_{e} {\bf u}_{e}$. Using \eqref{colvel}
\begin{equation}
{\bf j} = i \varepsilon_{0} \frac{\omega_{pe}^2}{\omega + i \nu_{ei}} \, {\bf E} ({\bf r}) e^{-i \omega t}= \sigma \, {\bf E} ({\bf r}, t),
\label{colcurrent}
\end{equation}
where the plasma frequency is defined as
\begin{equation}
\omega_{pe}^2 = n_{e} e^2/( \varepsilon_{0} m_{e} ),
\label{plasmafrequency}
\end{equation}
and the plasma conductivity $\sigma$ is a complex quantity,
\begin{equation}
\sigma = i \varepsilon_{0} \frac{\omega_{pe}^2}{\omega + i\nu_{ei}} .
\label{colconductivity}
\end{equation}

From Maxwell's equations for harmonic fields,
\begin{eqnarray}
\nabla \times {\bf E} ({\bf r}) &=& i \omega {\bf B} ({\bf r}), \nonumber \\
\nabla \times {\bf B} ({\bf r}) &=& \mu_{0} \sigma {\bf E} - i \frac{\omega}{c^2} {\bf E} ({\bf r}).
\label{maxwellstat1}
\end{eqnarray}
The second equation in (\ref{maxwellstat1}) can be written as
\begin{equation}
\nabla \times {\bf B} ({\bf r}) = - i \frac{\omega}{c^2} \varepsilon \, {\bf E} ({\bf r}),
\label{maxwellstat2}
\end{equation}
where
\begin{equation}
\varepsilon = 1 - \frac{\omega_{pe}^{2}}{\omega(\omega + i\nu_{ei})},
\label{dielectric}
\end{equation}
is the dielectric function of the plasma. Taking the curl in equations (\ref{maxwellstat1}) and (\ref{maxwellstat2}), one gets the wave equations
\begin{eqnarray}
\nabla^2 {\bf E} ({\bf r}) &-& \nabla \left( \nabla \cdot {\bf E} ({\bf r}) \right) + \frac{\omega^2}{c^2} \varepsilon \, {\bf E} ({\bf r}) = 0 , \nonumber \\
\nabla^2 {\bf B} ({\bf r}) &+& \frac{1}{\varepsilon} \nabla \varepsilon \times \left( \nabla \times {\bf B} ({\bf r}) \right) + \frac{\omega^2}{c^2} \varepsilon \, {\bf B} ({\bf r}) = 0 ,
\label{wave}
\end{eqnarray}
that give the spatial behaviour of the electric and magnetic fields in the plasma.

In the case of a neutral plasma with an uniform density $\nabla \varepsilon = 0$ and $\nabla \cdot {\bf E} = 0$, for harmonic electromagnetic waves ${\bf E} ({\bf r})\sim \exp (i {\bf k}\cdot{\bf r})$, the equations \eqref{wave} yield  $\omega^2 \varepsilon = c^2 k^2$.  So using \eqref{dielectric}
\begin{equation}
\omega^2 = \omega_{pe}^2 \left( 1 - i \frac{\nu_{ei}}{\omega} \right) + c^2 k^2 ,
\label{wavestatcol1}
\end{equation}
in which it is assumed that $\nu_{ei} \ll \omega$. Equation (\ref{wavestatcol1}) means that electromagnetic waves are damped. Writing $\omega = \omega_{r} - i \nu/2$, where $\nu$ is the damping rate, equation (\ref{wavestatcol1}) becomes
\begin{eqnarray}
\omega_{r} &=& \sqrt{\omega_{pe}^2 + k^2 c^2 }, \nonumber\\
\nu &=& \frac{\omega_{pe}^2}{\omega_{r}^2} \, \nu_{ei} .
\label{wavestatcol2}
\end{eqnarray}
The damping rate can be computed from the zero-order distribution function. This allows to get an expression for the collision frequency, namely \cite{ref19,spitzer}
\begin{equation}
\nu_{ei} = \frac{1}{3 (2 \pi )^{3/2}} \, \frac{Z \omega_{pe}^4}{n_{e} v_{e}^3} \, \ln{\Lambda}=3.61\times 10^{-6} Z\ln \Lambda \frac{n_e}{T^{3/2}},
\label{wavestatcol3}
\end{equation}
where we use $v_{e} = \sqrt{K_BT/m_{e}}$ for the thermal velocity of electrons. The factor $\Lambda$ is the ratio of the maximum and minimum impact parameters. The maximum impact parameter $r_{max}$ is given by the Debye length $\lambda_{De}$, as the Coulomb potential is shielded out over that distance. The minimum impact parameter is given by the classical distance of closest approach $r_{min}=Ze^2/4\pi\varepsilon_0 m_ev_e^2$, which averaging over all the particle velocities and assuming a Maxwellian distribution yields $\bar{r}_{min}=Ze^2/(12\pi\varepsilon_0 K_BT)$, so
\begin{equation}
\Lambda = \frac{\lambda_{De}}{\bar{r}_{min}}= \frac{12 \pi n_{e} \lambda_{De}^3}{Z}.
\label{wavestatcol4}
\end{equation}
For typical values of plasmas, $\ln \Lambda \approx 10$.

We must notice that \eqref{wavestatcol3} is only an approximation where the zero order electron distribution is taken Maxwellian. For non equilibrium processes, the collisional damping rate could be less. For  example in the case of a  super-Gaussian distribution, the collisional damping is reduced by a factor of 2. 

\section{The acceleration of electrons in atmospheric plasmas}
In this section we will study how electrons in a plasma are accelerated by an electromagnetic pulse propagating through the plasma. We will first set the equations to describe the electron dynamics and then compare their predictions with some particle in cell code simulations to verify the results. 

Let us consider an electromagnetic pulse which is propagating in the $x$-direction inside a plasma,
\begin{equation}
{\bf E}({\bf r},t)=\int dk A(k)e^{i(k x-\omega t)}\hat{\bf e}_y + \mathrm{c.c.}\, ,
\end{equation}
where $A(-k)=A(k)^*$ and $A(k)$ for $k>0$ is nonzero only in the vicinity of a central wavenumber $k_0$. From the relation \eqref{wavestatcol2} we take for the pulse a lead frequency $\omega_0=(w_{pe}^2+c^2k_0^2)^{1/2}$. Expanding around $k_0$ yields the following expressions for the electromagnetic field,
\begin{eqnarray}
{\bf E}({\bf r},t)=E_0 \, f(x-v_g t)\cos[(\omega_0-v_g k_0)t] \, \hat{\bf e}_y, \nonumber\\
{\bf B}({\bf r},t)=\frac{v_g}{c^2}E_0 \, f(x-v_g t)\cos[(\omega_0-v_g k_0)t] \, \hat{\bf e}_z,
\label{eq:pulse1}
\end{eqnarray}
where 
\begin{equation}
v_g=\left(\frac{\partial \omega}{\partial k}\right)_{k=k_0} = \frac{c^2 k_0}{\omega_0} = c \, \varepsilon^{1/2}
\end{equation}
is the propagation speed inside the plasma, $\varepsilon$ is the plasma dielectric function (\ref{dielectric}), and $f(x)=E_y({\bf r},0)/E_0$ gives the shape of the pulse, $E_0$ being the amplitude of the electric field. 
 
As discussed in the previous section, the condition $\omega_0 \ge \omega_{pe}$ is required for the propagation of the electromagnetic pulse inside the plasma. In the following we will consider $\omega_0 \ge 10 \, \omega_{pe}$. For $\omega_0 = 10 \, \omega_{pe}$, $v_g$ differs from the speed of light in vacuum in less than 1\%. For larger values of $\omega_{0}$, the pulse propagation speed is even closer to $c$. The electromagnetic pulse will start interacting with the electrons at $t=0$. The interaction will finish at a time of the order of $2\pi/\omega_0$. Since $\omega_0-v_gk_0=\omega_{pe}^2/\omega_0$, the cosine term in (\ref{eq:pulse1}), $\cos[(\omega_0-v_g k_0)t]=\cos[(\omega_{pe}/\omega_0)^2 \omega_0t]\approx 1$ can be safely neglected throughout the process. We will also assume that the damping rate (\ref{wavestatcol3}) of the pulse is negligible. Those assumptions will impose some constrains that we will discuss in the next section. With these approximations, the electromagnetic field can be written as
\begin{eqnarray}
{\bf E}({\bf r},t)=E_0 \, f(x-c t) \hat{\bf e}_y,\label{eq:pulse2a} \\
{\bf B}({\bf r},t)=\frac{E_{0}}{c} f(x-c t) \hat{\bf e}_z.
\label{eq:pulse2}
\end{eqnarray}

Since the pulse frequency is larger than the plasma frequency, the dynamics of the electrons will be collisionless while interacting with the pulse, so can be modelled by 
\begin{equation}
\frac{d{\bf p}}{dt}=-e{\bf E}-e{\bf v}\times{\bf B},
\label{eq:eom1}
\end{equation}
where ${\bf p}=m_e\gamma{\bf v}$ is the relativistic momentum of the electron and $\gamma=(1-v^2/c^2)^{-1/2}$. Equation (\ref{eq:eom1}) can be written as \cite{landau}
\begin{equation}
\frac{d{\bf v}}{dt}=-\frac{e}{m_e}\left(1-\frac{v^2}{c^2}\right)^{1/2}\left[{\bf E}+{\bf v}\times{\bf B}-\frac{{\bf v}({\bf v}\cdot{\bf E})}{c^2}\right].
\label{eq:eom2}
\end{equation}
We define dimensionless variables $\tilde{t}=t e E_0/(m_e c)$ and $\tilde{\bf r}={\bf r}e E_0/(m_e c^2)$, so that the scaled velocity is just $\tilde{\bf v}={\bf v}/c$. Using (\ref{eq:pulse2a}) and \eqref{eq:pulse2}, the equation of motion (\ref{eq:eom2}) becomes
\begin{eqnarray}
\frac{d\tilde{v}_x}{d\tilde{t}}=-\sqrt{1-\tilde{v}^2}(\tilde{v}_y-\tilde{v}_x\tilde{v}_y)\, \tilde{f}[\tilde{\omega}(\tilde{x}-\tilde{t})], \nonumber\\
\frac{d\tilde{v}_y}{d\tilde{t}}=-\sqrt{1-\tilde{v}^2}(1-\tilde{v}_y^2-\tilde{v}_x)\, \tilde{f}[\tilde{\omega}(\tilde{x}-\tilde{t})], \nonumber\\
\frac{d\tilde{v}_z}{d\tilde{t}}=-\sqrt{1-\tilde{v}^2}(-\tilde{v}_z\tilde{v}_y)\, \tilde{f}[\tilde{\omega}(\tilde{x}-\tilde{t})], 
\label{eq:eom3}
\end{eqnarray}
where $\tilde{f}(k_0 x)=f(x)$ takes into account the assumed form of the wave-packet, and 
\begin{equation}
\tilde{\omega}=\frac{m_e c\,\omega_0}{e E_0}.
\label{eq:omtilde}
\end{equation} 
We are interested in pulses which are able to accelerate electrons to kinetic energies $K=m_ec^2[(1-\tilde{v}^2)^{-1/2}-1]$ in the order of several MeV after the interaction time, which is $\tilde T\approx2\pi/\tilde{\omega}$. Therefore, in this model the interaction with the pulse is controlled by the dimensionless parameter $\tilde{\omega}$. Pulses with different amplitudes and frequencies but same ratio $\omega_0/E_0$ (and shape) produce the same acceleration.

\begin{figure}
\begin{center}
\includegraphics[width=0.5\textwidth]{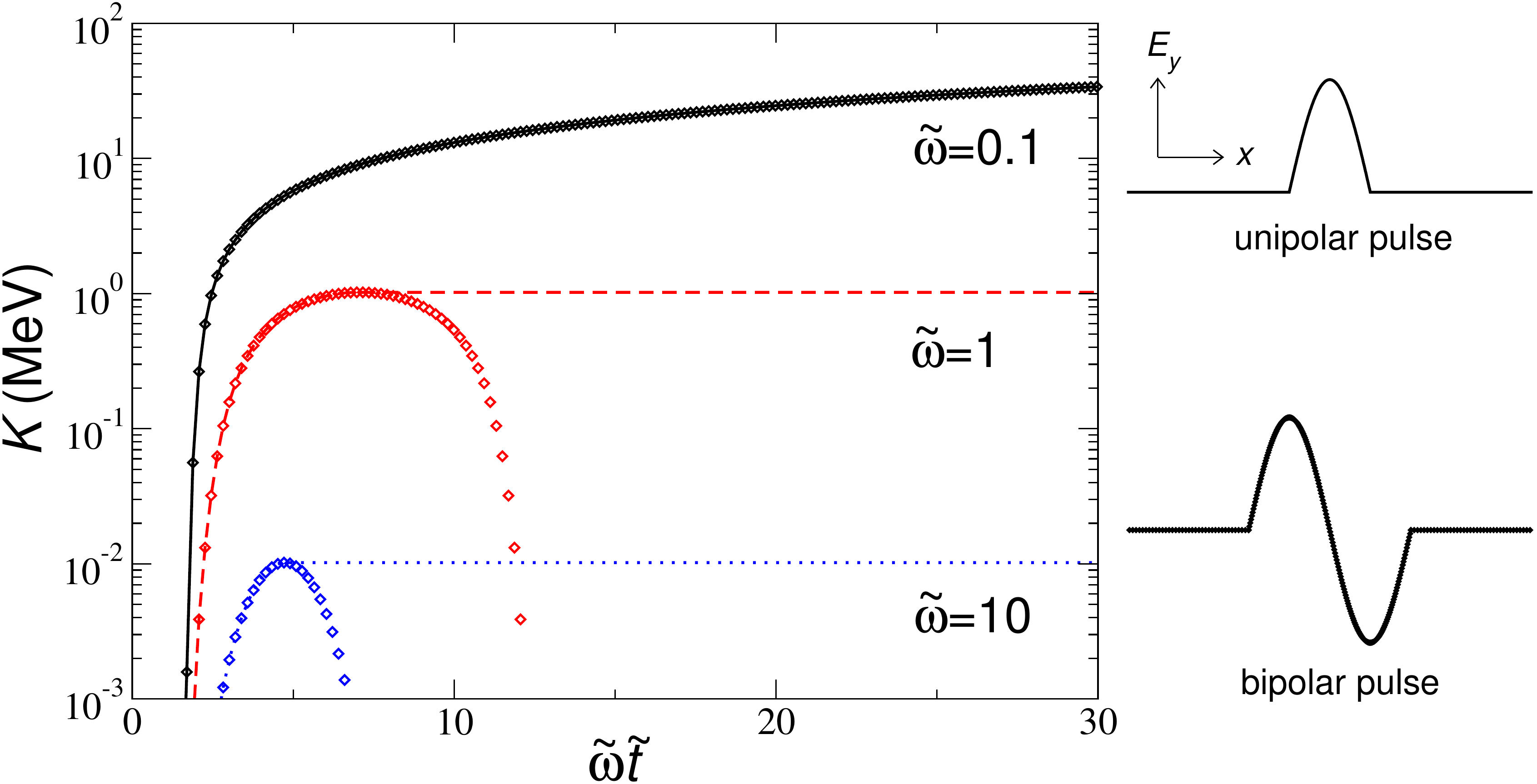}
\end{center}
\caption{Kinetic energy vs time of an electron accelerated by a unipolar (dotted lines) and a bipolar (crosses) EM pulse, for
several $\tilde{\omega}$. The pulses, depicted in the right side, are assumed to be translation invariant in the directions perpendicular to the direction of propagation (along the $x$-axis).}
\label{fig:K_t} 
\end{figure}

Figure \ref{fig:K_t} shows the relativistic kinetic energy obtained for a free electron initially at rest at the origin accelerated by a pulse with the shape of the form
\begin{equation}
\tilde{f}(\tilde{\omega}x)=\cos[\tilde{\omega}(x-\tilde{T}/4)][H(x+\tilde{T}/2)-H(x)],
\label{eq:fx}
\end{equation}
which is unipolar (see right of Fig. \ref{fig:K_t}), and also for the bipolar shape
\begin{equation}
\tilde{f}(\tilde{\omega}x)=\cos[\tilde{\omega}(x-\tilde{T}/4)][H(x+\tilde{T})-H(x)],
\label{eq:fx2}
\end{equation}
where $H(x)$ is the Heaviside step function. The results were obtained by solving numerically (\ref{eq:eom3}). The simulations show that for values $\tilde{\omega} \lesssim 0.1$, regardless whether the shape is unipolar or bipolar, the kinetic energy keeps growing after five periods $T_0$, which is an indication that the electron is actually trapped by the pulse, soon achieving very high energies. For  $\tilde{\omega}\approx 1$, the electromagnetic pulse is not so strong as to carry away the electron. It is still able to produce accelerations to energies in the MeV scale, though for the bipolar pulse the electron acceleration is later reversed by the own electromagnetic packet. For larger values of the reduced frequency, $\tilde{\omega}  \gtrsim 10$, the impulse produced by the pulse is well below the MeV range. 

\begin{figure}
\begin{center}
\includegraphics[width=0.5\textwidth]{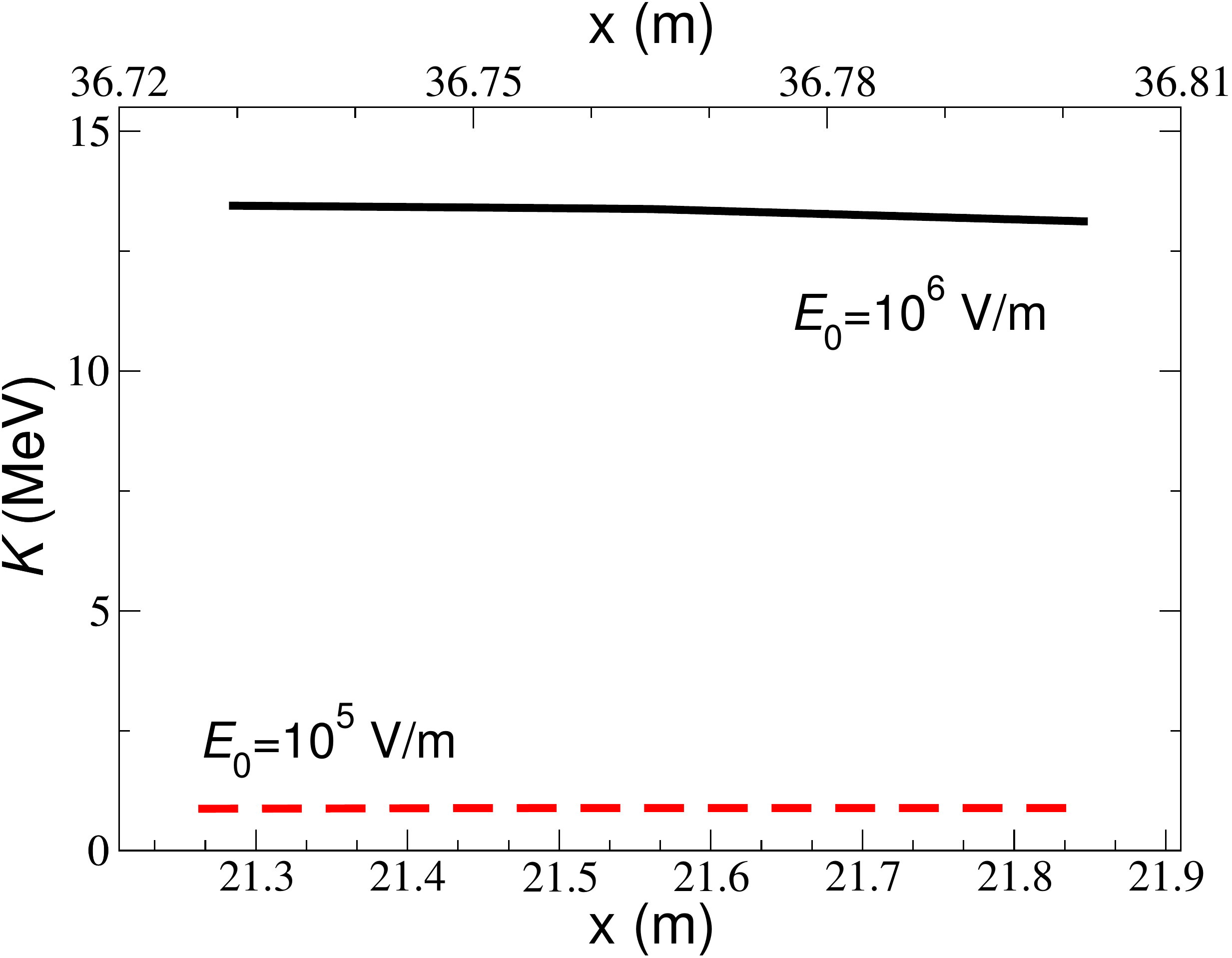}
\end{center}
\caption{Energy of electrons in a plasma of density $n_0=10^{10} \, \mbox{m}^{-3}$ after an acceleration time interval of $1.44\times 10^{-7} \, \mbox{s}$ due to an electromagnetic pulse $\omega_0 = 2 \pi \times 10^7 \, \mbox{rad}/\mbox{s}$. The continuous line corresponds to a unipolar pulse of amplitude $E_0 = 10^6 \, \mbox{V}/\mbox{m}$ and its abscissa is the upper one.  The dashed line correspond to $E_0 = 10^5 \, \mbox{V}/\mbox{m}$, being its abscissa the lower one. The plasma frequency is $\omega_{pe} = 5.6 \times 10^6 \, \mbox{rad}/\mbox{s}$.}
\label{fig:vorpal} 
\end{figure}

These results were numerically verified using the Finite Difference Time Domain Particle in Cell software VORPAL \cite{vorpal}. Figure \ref{fig:vorpal} shows the kinetic energy profile after a time interval of $1.44 \times 10^{-7} \, \mbox{s}$, for a plasma of density $n_0 = 10^{10} \, \mbox{m}^{-3}$ with a unipolar pulse defined by the boundary condition ${\bf E}(0,y,z,t) = E_0\, \sin(\omega_0 t) \, H(\pi/\omega_0-t)\, \hat{\bf e}_y$, where $H$ is the Heaviside step function. For $E_0 = 10^5 \, \mbox{V}/\mbox{m}$ and $\tilde{\omega}=1.071$, as predicted, the pulse is able to produce electrons just in the MeV scale, whereas the pulse with $E_0 = 10^6 \, \mbox{V}/\mbox{m}$ yields $\tilde{\omega}=0.1071$, able to accelerate free electrons to very large energies. Note that for the case of the higher electric field amplitude, the acceleration occurs at larger distances. 

\section{A proposed scenario for Gamma flashes in planetary atmospheres}
We have shown in previous section how an electromagnetic pulse, interacting with a created plasma, is able to accelerate electrons to high MeV energies provided some conditions are fulfilled. We propose the following scenario to this process takes place in planetary atmospheres. 

\begin{figure}
\begin{center}
\includegraphics[width=0.5\textwidth]{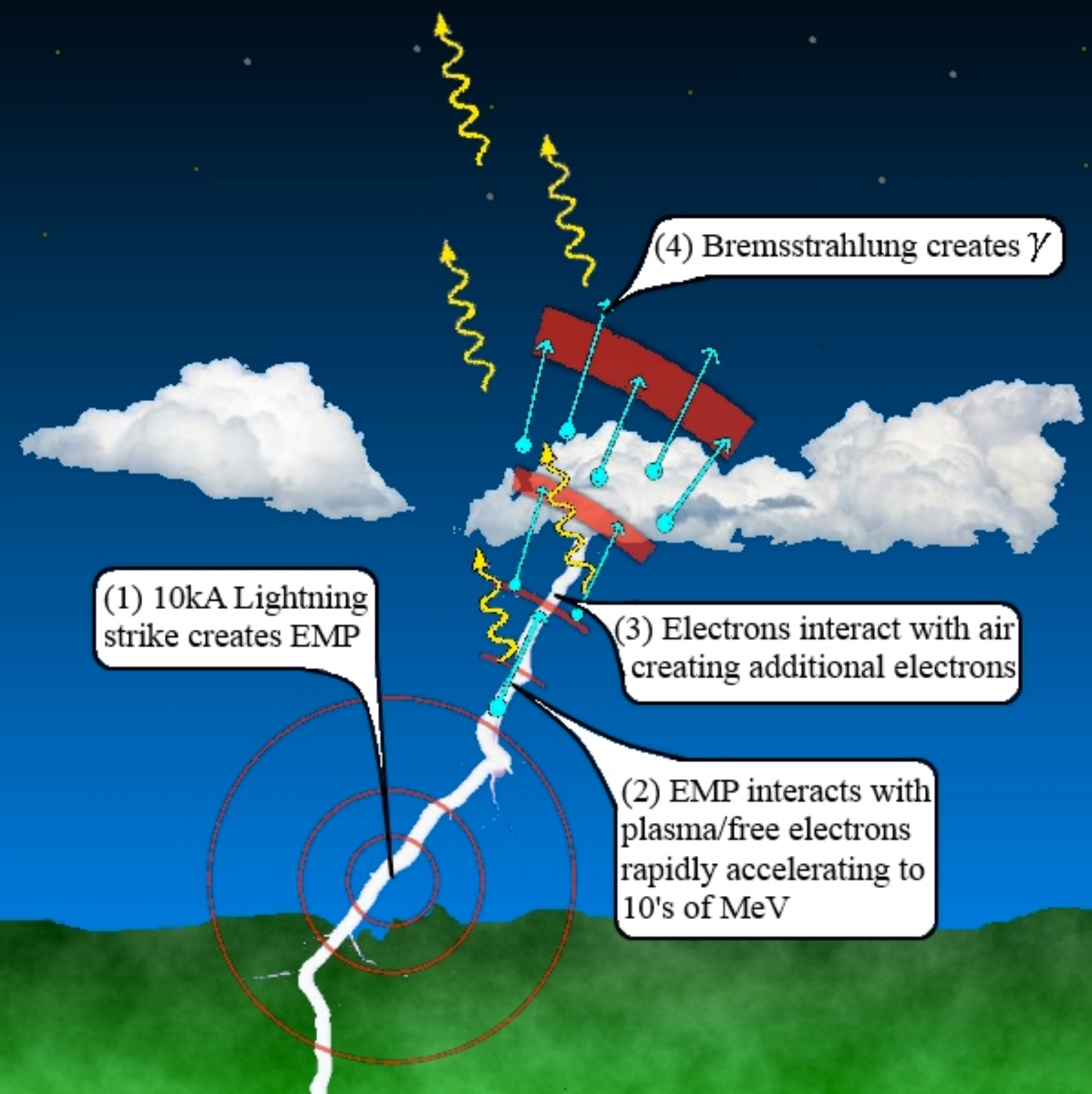}
\end{center}
\caption{Cartoon outlining the events leading to the creation of a TGF, courtesy of D. Burton.}
\label{fig1} 
\end{figure}

Some electromagnetic activity has to produce the pulse and the plasma, for example lightning discharges cloud to cloud, or cloud to earth. When the pulse is created, it propagates at basically the speed of light through the planetary atmosphere. If that pulse finds in its way a plasma, then the wakefield mechanism might be fired. We can speculate that the origin of the plasma is another discharge, or even the same discharge that creates the pulse. We will come back to this point in the next section when studying the Earth atmosphere case. The high energy electrons accelerated by the wakefield, leaving the plasma region, can interact with neutral molecules, creating an additional shower of secondary electrons, other particles and gamma radiation via Bremsstrahlung. In Figure \ref{fig1}, a cartoon picture summarises the whole cascade of events, leading eventually to the creation of a gamma flash. 

From previous sections we need the following conditions for this scenario to work:
\begin{itemize}
\item The numbers of particles in the Debye sphere must be large,
\begin{equation}
 N_D  \gg 1\label{28}.
\end{equation}
\item The Debye length must be much smaller than the characteristic size of the plasma region, 
 \begin{equation}
 \lambda_{De} \ll L. \label{29}
 \end{equation}
\item The frequency of electron and ions collisions must be smaller than the electromagnetic pulse angular frequency,
 \begin{equation}
 \nu_{ei}\ll \omega_0.\label{30}
 \end{equation}
\item The damping rate must be negligible,
\begin{equation}
\nu=\frac{\omega_{pe}^2}{\omega_0^2}\nu_{ie} \ll 1.\label{33}
\end{equation}
\item For the acceleration of electrons to MeV energies, we need the condition
\begin{equation}
\tilde{\omega}=\frac{m_e c \omega_0}{eE_0}< 1.\label{34}
\end{equation}
\end{itemize}


\section{Predictions for TGF's}
 In this section we consider the ignition of TGF's in the Earth atmosphere by the wakefield mechanism. The occurrence of transient luminous events (TLE's) are candidates for the electromagnetic pulse source.  At the troposphere altitudes, from 0 to 20 km, we have conventional cloud to ground lightning discharges of both types, positive and negative ones, and cloud to cloud discharges. Closer to the stratosphere, from 20 to 40 km, there are blue jets \cite{Wescott}. At higher altitudes, at the mesosphere, we find sprites and, at the ionosphere, elves \cite{EarleWilliams}.  

Due to the fact that it seems that TGF's appear at 15 km as pointed at the introduction, we will assume that the electromagnetic pulse is produced by lightning activity.  We will take for the peak intensity of the pulse $5\times10^5$ V/m \cite{Uman1}. From relation \eqref{34}, we can get the condition $\omega_0<3\times 10^8$ rad/s. According to \cite{uman,willet}, the range of average frequency spectra of return strokes is $0.2 - 20$ MHz, so we can have a pulse with $f_0=\omega_0/2\pi<5\times 10^7$ Hz. 

To fulfill \eqref{30} and \eqref{33}, as in the simulations, we take $\omega_{pe}=\omega_0/10$. It turns out that there is a critical maximum electron density of the plasma $n_{ec}\approx 3\times 10^{11}\,\, \mbox{m}^{-3}$ which allows wakefield acceleration. With those values, $\ln \Lambda=14$, where we have taken the mean atomic number of the air $Z=7.2$, and the electrom plasma temperature $10^4\,\,\mbox{K}$ \cite{ref24}. The characteristic plasma length is $\lambda_{De}=1$ cm and  \eqref{30} $\nu_{ei}=100$ rad/s which is much less than $\omega_0$ so collisions can be negleted.

Can a plasma of of size bigger than 1 cm and electron density less than  $3\times 10^{11}\,\, \mbox{m}^{-3}$ be created? At altitudes around 15 km, the air density in the Earth is bigger than $3\times 10^{11}\,\, \mbox{m}^{-3}$, so the total ionisation of the air will give a too dense plasma. However, the rapid expansion of a plasma, due to the shock wave created by a lightning discharge could account for such a low  density plasma. At higher altitudes, the plasma density can be lower, but the electromagnetic pulse normally decays inversely with the distance as $1/r$. Then if the electromagnetic pulse is created at large distances from the plasma region, the field will not be able to trigger the wakefield acceleration when reaching the plasma. We have considered only the fields created by cloud to ground dischages in the discussion, but there are other sources of TLE's which might provided the necessary pulses. At the present state of experimental data, we cannot go further in this discussion.

\section{Further discussions and conclusions}
In this paper we have studied the conditions for the creation of MeV electrons in atmospherics plasmas. An intense electromagnetic pulse interacting with the plasma can create a wake on the plasma. Electrons trapped in such oscillations can be accelerated under certain conditions to high energies. We have shown that those electrons could reach energies in the MeV range, thus being able to ignite gamma bursts. We propose a scenario for the wakefield acceleration to take place in planetary atmospheres, where the pulse and the plasma are generated by electromagnetic activity. 

Applying our theory to the origin of MeV or runaway electrons for the production of Terrestrial Gamma Flashes (TGF's) we predict that an electromagnetic pulse of  $5\times 10^5$ V/m peak can account for the ignition of TGF's provided that the pulse encounters a plasma of low electron density. 

The idea of TGF's production by an electromagnetic pulse from a lightning return stroke propagating into the air was explored previously \cite{lehtinen}. The conclusion was that very high electromagnetic field amplitudes are necessary to overcome the friction due to collisions with the nonionized gas.      

The novelty of our proposal is the interaction of the pulse with a plasma already present in the atmosphere and created for example by another discharge or any electromagnetic activity. In order to accelerate electrons to MeV energies, the amplitudes of the pulses depend on the density of the plasma. Under the conditions studied in this work, the wakefield mechanism trigger the ignition of TGF's. 
We stress that if those conditions are fulfilled, the pulse can propagate inside the plasma without damping and electrons will be accelerated dragged by the pulse. 

In order to address and clarify the situation, there is scheduled the Atmosphere-Space Interactions Monitor (ASIM ESR) mission on the ISS external facilities of the Columbus module. It will be an Earth observation facility devoted to the study of severe thunderstorm and their role in Earth's climate. Among other activities, ASIM will study giant electrical discharges (lightning) in the high-altitude atmosphere above thunderstorms, and the precise location of TGF's \cite{web}.

\section*{Acknowledgement}
 This work has been partially supported by the Spanish Ministerio de Econom\'{\i}a y Competitividad, under project ESP2013-48032-C5-2-R.


\section*{References}
\begin{thebibliography}{99}
\bibitem{dawson} T. Tajima and J. Dawson, {\it Phys. Rev. Lett.} {\bf 43}, 267 (1979).
\bibitem{ref1} G. J. Fishman et al., {\it Science} {\bf 264}, 1313 (1994).
\bibitem{web} https://directory.eoportal.org/web/eoportal/satellite-missions/i/iss-asim.
\bibitem{ref9} T. Gjesteland et al., {\it J. Geophys. Res.} {\bf 115}, A00E21 (2009).
\bibitem{ref10} B. Grefenstette et al., {\it J. Geophys. Res.} {\bf 114}, A02314 (2009).
\bibitem{ref11} G. J. Fishman et al., {\it J. Geophys. Res} {\bf 116}, A07304 (2011).
\bibitem{ref2} D. M. Smith et al.,  {\it Science} {\bf 307} (5712), 1085 (2005). 
\bibitem{ref2p}S. A. Cummer et al., {\it Geophys. Res. Lett.} {\bf 32}, L08811 (2005).
\bibitem{ref3} M. Marisaldi et al., {\it J. Geophys. Res.} {\bf 115}, A00E13 (2010).
\bibitem{ref12} M. Tavani et al., {\it Phys. Rev. Lett.} {\bf 106}, 018501 (2011).
\bibitem{ref17} B. Hazelton et al., {\it Geophys. Res. Lett.} {\bf 36}, L01108 (2009).
\bibitem{ref18} T. Gjesteland et al., {\it J. Geophys. Res.} {\bf 116}, A11313 (2011).
\bibitem{ref5} U. S. Inan et al., {\it Geophys. Res. Lett.} {\bf 23}, 1017 (1996).
\bibitem{ref6} J. R. Dwyer and D. M. Smith, {\it Geophys. Res. Lett.} {\bf 32}, L22804 (2005).
\bibitem{armenia} A. Chilingarian, {\it Journal of Atmospheric and Solar-Terrestrial Physics} {\bf 107}, 68 (2014).
\bibitem{ref7} M. A. Stanley et al., {\it Geophys. Res. Lett.} {\bf 33}, L06803 (2006).
\bibitem{ref8} X. M. Shao, T. Hamlin and D. M. Smith,  {\it J. Geophys. Res.} {\bf 115}, A00E30 (2010).
\bibitem{ref13} A. V. Gurevich, G. M. Milikh and R. Roussel-Dupr\'e, {\it Phys. Lett. A} {\bf 165}, 463 (1992).
\bibitem{ref15} S. Celestin and V. P. Pasko, {\it J. Geophys. Res.} {\bf 116}, A0605 (2011).
\bibitem{ref16} B. E. Carlson et al., {\it J. Geophys. Res.} {\bf 114}, A00E08 (2012).
\bibitem{lehtinen} U. S. Inan and N. G. Lehtinen, {\it Geophys. Res. Lett.} {\bf 32}, L19818 (2005).
\bibitem{ref19} W. L. Kruer, {\it The Physics of Laser Plasma Interactions} (Westview Press, 2003)
\bibitem{spitzer} L. Spitzer Jr. {\it Physics of Fully Ionized Gases} (John Wiley \& Sons, 1962). 
\bibitem{landau} L. D. Landau and E. M. Lifshitz, {\it The Classical Theory of Fields} (Elsevier, 1975), pp. 52.
\bibitem{vorpal} C. Nieter and J. R. Cary, {\it J. Comp. Phys.} {\bf 196}, 448 (2004).
\bibitem{Wescott} E. M. Wescott et al., {\it Geophys. Res. Lett.} {\bf 23}, 2153 (1996).
\bibitem{EarleWilliams} E. Williams, {\it Physics Today}, November 2001, 41-47.
\bibitem{Uman1} M. A. Uman, {\it J. Geophys. Res.} {\bf 90}, 6121 (1985).
\bibitem{uman} V. A. Rakov and M. A. Uman, {\it Lightning Physics and Effects} (Cambridge University Press, 2003), pp. 158-159.
\bibitem{willet} J. C. Willet et. al., {\it J. Geophys. Res.} {\bf 95}, 20367 (1990).
\bibitem{ref24} N. Liu and V. P. Pasko, {\it J. Geophys. Res.} {\bf 109}, A04301 (2004).




\end{thebibliography}
\end{document}